\documentclass[pra,aps,twocolumn,superscriptaddress]{revtex4-2}
\usepackage{graphicx,empheq,amssymb,scalerel,tikz}
\usepackage[colorlinks=true, citecolor=blue, urlcolor=blue, linkcolor=blue ]{hyperref}
\newcommand{\ket}[1]{\left|#1\right\rangle}
\usetikzlibrary{svg.path}
\definecolor{orcidlogocol}{HTML}{A6CE39}
\tikzset{orcidlogo/.pic={
		\fill[orcidlogocol] svg{M256,128c0,70.7-57.3,128-128,128C57.3,256,0,198.7,0,128C0,57.3,57.3,0,128,0C198.7,0,256,57.3,256,128z};
		\fill[white] svg{M86.3,186.2H70.9V79.1h15.4v48.4V186.2z}
		svg{M108.9,79.1h41.6c39.6,0,57,28.3,57,53.6c0,27.5-21.5,53.6-56.8,53.6h-41.8V79.1z M124.3,172.4h24.5c34.9,0,42.9-26.5,42.9-39.7c0-21.5-13.7-39.7-43.7-39.7h-23.7V172.4z}
		svg{M88.7,56.8c0,5.5-4.5,10.1-10.1,10.1c-5.6,0-10.1-4.6-10.1-10.1c0-5.6,4.5-10.1,10.1-10.1C84.2,46.7,88.7,51.3,88.7,56.8z};}}
\newcommand\orcid[1]{\href{https://orcid.org/#1}{\mbox{\scalerel*{\begin{tikzpicture}[yscale=-1,transform shape]\pic{orcidlogo};\end{tikzpicture}}{|}}}}

\begin{document}
\title{Majorana modes in trapped-ion system and their Floquet engineering}
\author{Ming-Jian Gao\orcid{0000-0002-6128-8381}}
\affiliation{School of Physical Science and Technology \& Lanzhou Center for Theoretical Physics, Lanzhou University, Lanzhou 730000, China}
\affiliation{Key Laboratory of Quantum Theory and Applications of MoE \& Key Laboratory of Theoretical Physics of Gansu Province, Lanzhou University, Lanzhou 730000, China}
\author{Yu-Peng Ma}
\affiliation{School of Physical Science and Technology \& Lanzhou Center for Theoretical Physics, Lanzhou University, Lanzhou 730000, China}
\affiliation{Key Laboratory of Quantum Theory and Applications of MoE \& Key Laboratory of Theoretical Physics of Gansu Province, Lanzhou University, Lanzhou 730000, China}
\author{Jun-Hong An\orcid{0000-0002-3475-0729}}
\email{anjhong@lzu.edu.cn}
\affiliation{School of Physical Science and Technology \& Lanzhou Center for Theoretical Physics, Lanzhou University, Lanzhou 730000, China}
\affiliation{Key Laboratory of Quantum Theory and Applications of MoE \& Key Laboratory of Theoretical Physics of Gansu Province, Lanzhou University, Lanzhou 730000, China}
\begin{abstract}
Obeying non-Abelian statistics, Majorana fermions holds a promise to implement fault-tolerant quantum computing. It was found that Majorana fermions can be simulated by the zero-energy excitation in a nanowire with strong spin-orbit coupling interacting with an $s$-wave superconductor under a magnetic field. However, the signal of Majorana fermion in that system is obscured by the disorder in the nanowire and the confinement potential at the wire end. Thus, more controllable platforms  are desired to simulate Majorana fermions. We here propose an alternative scheme to simulate the Majorana fermions in a trapped-ion system. Our dimerized-ion configuration permits us to generate the Majorana modes not only at zero energy but also at the nonzero ones, which enlarge the family of Majorana modes and supply another qubit carrier for quantum computing. We also investigate the controllability of the Majorana modes by Floquet engineering. It is found that a widely tunable number of Majorana modes are created on demand by applying a periodic driving on the trapped-ion system. Enriching the platforms for simulating Majorana fermions, our result would open another avenue for realizing fault-tolerant quantum computing.
\end{abstract}
\maketitle

\section{Introduction}
As a rapidly developing field in modern physics, topological phases not only enrich the paradigm of condensed matter physics, but also inspire many important applications in quantum technology \cite{RevModPhys.82.3045, RevModPhys.83.1057,RevModPhys.87.137,RevModPhys.88.035005, RevModPhys.90.015001, RevModPhys.93.025002}. Simulating the elusive Majorana fermions in particle physics \cite{PhysRevLett.100.096407,RN198,PhysRevLett.105.077001,Sato_2017,PhysRevLett.123.126804,PhysRevLett.131.086601,PhysRevLett.131.056001,PhysRevLett.130.156002,PhysRevLett.125.097001,PhysRevLett.123.177001,PhysRevLett.123.156801,PhysRevLett.122.236401,PhysRevLett.121.096803}, topological superconductor has become an ideal candidate to realize fault-tolerant quantum computing due to its unique non-Abelian statistics  \cite{PhysRevLett.86.268,RevModPhys.80.1083,PhysRevB.95.235305,doi:10.1073/pnas.1810003115,PhysRevB.99.195137,RN207,RN208}. It was theoretically found that the Majorana fermions can be simulated by the zero-energy excitation mode in a semiconducting nanowire with strong spin-orbit coupling interacting with an $s$-wave superconductor under a magnetic field \cite{PhysRevLett.105.077001}. Although having been realized \cite{doi:10.1126/science.1259327,RN203,doi:10.1126/science.aan4596,doi:10.1126/sciadv.aar5251,RN201,RN204,doi:10.1126/sciadv.abl4432,RN205,PhysRevLett.130.266002,Soldini2023}, the generation of Majorana mode in the nanowire systems is greatly hindered by the multiple subbands, the disorder in the nanowire, and the confinement potential at the nanowire's ends \cite{PhysRevLett.109.267002,PhysRevB.86.100503,PhysRevB.84.144522,PhysRevLett.107.196804}. Furthermore, on-demand generation and annihilation of different numbers of Majorana modes are a prerequisite for performing quantum computing by braiding the Majorana modes. However, the number of Majorana modes simulated in the nanowire systems is hard to change anymore once the material sample is fabricated. Therefore, more platforms with better controllability to simulate the Majorana fermions are highly desired. 

A trapped-ion system has been widely used in quantum simulation \cite{RN226,RevModPhys.93.025001,RN234}. Its precise controllability makes it capable of simulating the behavior of many complicated systems, such as spin interactions \cite{doi:10.1126/science.1251422,PhysRevX.5.021026}, many-body localization \cite{doi:10.1126/science.aau4963,RN227}, prethermalization \cite{doi:10.1126/sciadv.1700672}, and nonequilibrium phases \cite{RN228,PhysRevLett.118.030401,RN229,PhysRevLett.119.080501}. However, the scheme on simulating the Majorana fermions in trapped-ion systems is still rare. On the other hand, coherent control via periodic driving of external fields called Floquet engineering offers an attractive control dimension to manipulate Majorana modes. Many novel topological phases have been created by periodic driving \cite{PhysRevB.87.201109,PhysRevA.100.023622,PhysRevB.102.041119,PhysRevB.103.L041115,PhysRevLett.121.036401,PhysRevLett.123.016806,PhysRevLett.124.057001,PhysRevLett.124.216601,PhysRevB.103.L041115,PhysRevB.103.115308,PhysRevB.104.205117,RN209,PhysRevResearch.2.013124,PhysRevResearch.3.023039,PhysRevLett.121.076802}. The well-established laser-control technique gives the trapped-ion system a natural advantage to realize Floquet engineering \cite{Zhang2022}. These advances show that the trapped-ion system has the advantage to become a more controllable platform than the nanowire systems to simulate the Majorana fermions. 

\begin{figure}[tbp]
\centering
\includegraphics[width=\columnwidth]{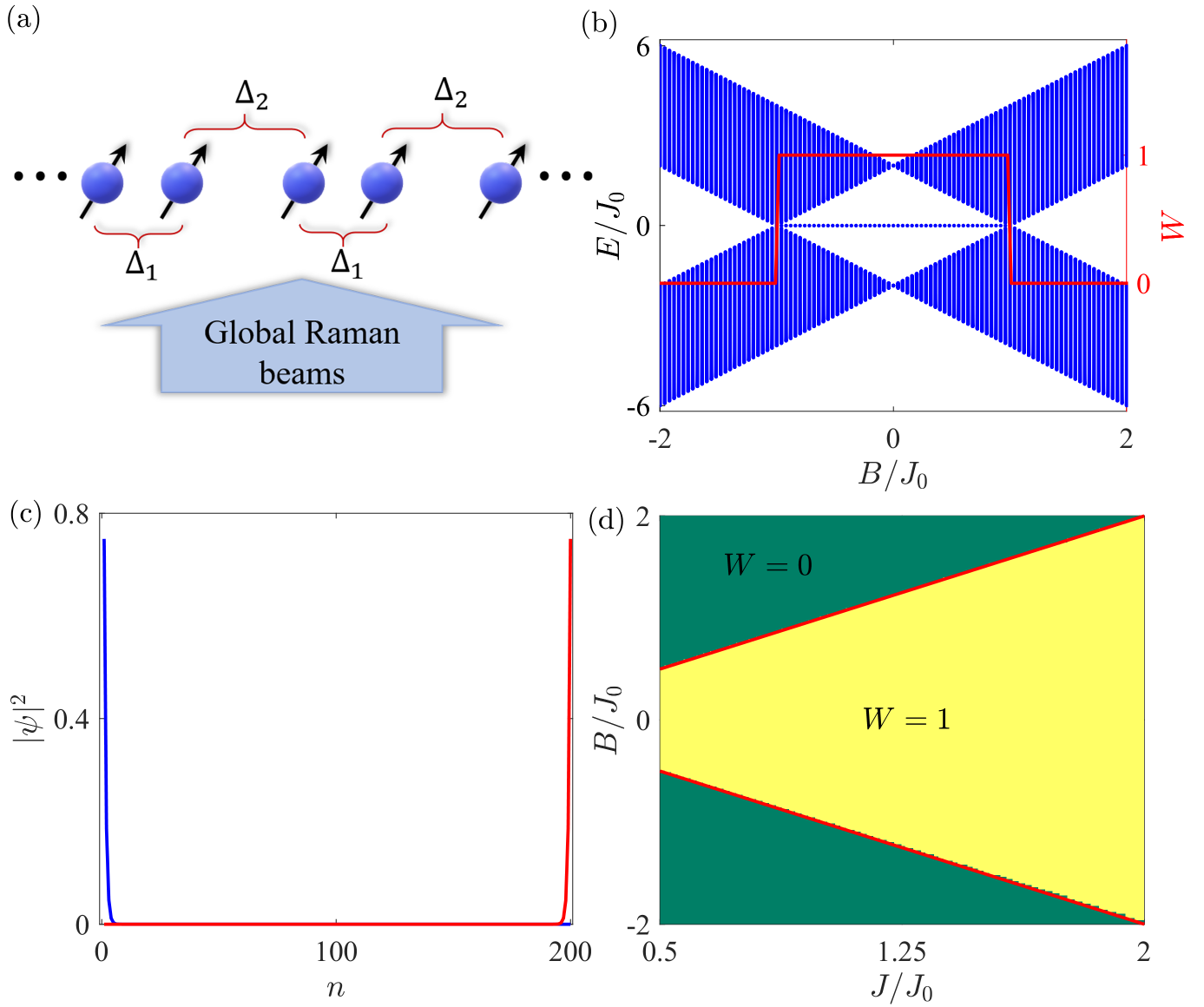}
\caption{(a) Schematic diagram of our dimerized trapped-ion system. (b) Energy spectrum of Eq. \eqref{SC} and winding number $W$ (red line) as a function of $B$ when $J_1=J_2=J_0$. (c) Probability distribution of the two zero-energy Majorana modes. (d) Phase diagram described by $W$.}\label{fig:1}
\end{figure}
We here propose a scheme to simulate the Majorana fermions in the trapped-ion system. We discover the formation of Majorana modes in this system. A spatial dimerization configuration of the ion separation is discovered to support the Majorana modes not only at zero energy but also at the nonzero ones. A complete topological description to these phases is established in the system. On the one hand, these rich phases enlarge the family of the Majorana modes. On the other hand, they offers more candidates to encode information for quantum computing. We then propose to control the Majorana modes of our system by Floquet engineering. A widely tunable number of Majorana modes are generated by the periodic driving in the regions where the static system does not host their existence. Our work enriches the platform for simulating the Majorana fermions and provides a feasible way to freely control different numbers of Majorana modes. It lays a foundation on exploring the realization of fault-tolerant quantum computing by the braiding of Majorana modes in trapped-ion systems.

\section{Majorana modes in trapped-ion systems}
Our scheme to simulate Majorana fermions in trapped-ion system is based on the idea to implement the transverse-field Ising model by globally applying two noncopropagating Raman laser beams on the ions. We consider a system consisting of a chain of $N$ $^{173}$Yb$^+$ ions confined in a linear Paul trap [see Fig \ref{fig:1}(a)]. Each ion has two hyperfine ‘‘clock'' states $|F=0,m_F=0\rangle\equiv \ket{\downarrow_z}$ and $\ket{F=1,m_F=0} \equiv \ket{\uparrow_z}$ of the $^2S_{1/2}$ valence electron and spin-1/2 nucleus, which act as the two orthogonal states of a pseudospin-1/2 system \cite{RN197,RN194,RN195,RN196,10.1063/1.5088164}. The two states have a frequency splitting $\omega_0=12.6\times2\pi$ GHz. Each ion is prepared in $\ket{\downarrow_x}=(\ket{\downarrow_z}-\ket{\uparrow_z}/\sqrt{2}$ by applying a laser pulse. Then, two noncopropagating Raman laser beams, with bichromatic beatnotes at frequencies $\omega_0\pm\mu$ and wave vector difference $\delta k$ pointing along the $x$ direction, are uniformly applied on the ions to generate a spin-dependent force at frequency $\mu$ on the ions. Under the rotating-wave approximation and in the Lamb-Dicke limit, we obtain an interaction Hamiltonian ($\hbar=1$)
\begin{equation}\hat{H}(t)=\Omega\sin(\mu t)\sum_{j=1}^N\sum_m\eta_{j,m}(\hat{a}_me^{-i\omega_mt}+\text{H.c.})\hat{\sigma}_j^x,\label{ham1}
\end{equation} 
where $\Omega$ is the Rabi frequency of the two Raman laser beams, $\hat{a}_m$ is the annihilation operator of the $m$th mode of the phonon, and $\eta_{j,m}=\delta k b_{j,m}/\sqrt{2M\omega_m}$, with $b_{j,m}$ being the normal-mode transformation matrix of the $j$th ion in the $m$th normal mode \cite{PhysRevLett.103.120502}. When the optical beatnote frequency is far detuning from one of each normal mode, the phonons are only virtually excited and the ion displacements become negligible. In this case, a Magnus expansion \cite{PhysRevLett.91.187902,PhysRevLett.97.050505} to the evolution operator of Eq. \eqref{ham1} results in $\hat{U}(t)=\exp(-it\sum_{i,j}J_{i,j}\hat{\sigma}_i^x\hat{\sigma}_j^x)$ with $J_{i,j}=\Omega^2{(\delta k)^2\over 2M}\sum_m{b_{i,m}b_{j,m}\over \mu^2-\omega_m^2}$, see Appendix \ref{appsys}. Further, adjusting the two Raman beatnotes to $\omega_0\pm\mu+B$, a uniform effective transverse magnetic field of $B$ along $\hat{\sigma}_i^z$ is generated \cite{RN194}. Thus, the dynamics of a transverse-field Ising Hamiltonian $\hat{ H}=\sum_{i,j}J_{i,j}\hat{\sigma}_i^x\hat{\sigma}_j^x+B\sum_i\hat{\sigma}_i^z$ is simulated in the trapped-ion system. The coupling strength $J_{ij}$ is approximated as a power law $J_{i,j}\simeq J_0/|z_i-z_j|^\beta$, with $\beta\in (0,3)$ \cite{RN197,RN194,RN195,doi:10.1126/science.1232296,PhysRevLett.92.207901}. We set $\beta=3$ by tuning the detuning between the beatnote frequency $\mu$ and the sideband $\omega_m$ \cite{RN194,RN195}. To generate the Majorana modes, we propose that the ion array has a spatial dimerization configuration \cite{PhysRevB.90.014505}, which can be realized by setting the distance between each odd (even) ion and its next neighboring ion being $\Delta_1$ ($\Delta_2$). Thus, the ion array reduces into a lattice of $N/2$ unit cells, each of which contains two sublattices labeled by $a$ and $b$. Keeping only the nearest-neighbor hopping of the dimerized ion array and making the Jordan-Wigner transformation \cite{PhysRevLett.115.177204}, we obtain the fermionized Hamiltonian, see Appendix \ref{appsys},
\begin{eqnarray} 
\hat{H}&=&\sum_{l=1}^{N/2-1}[J_1\hat{c}_{a,l}^{\dagger}(\hat{c}_{b,l}+\hat{c}_{b,l}^{\dagger})+J_2\hat{c}_{b,l}^{\dagger}(\hat{c}_{a,l+1}+\hat{c}_{a,l+1}^{\dagger})\nonumber\\
&&+\text{h.c.}]-2B\sum_{j=a,b}\sum_{l=1}^{N/2}\hat{c}_{j,l}^{\dagger}\hat{c}_{j,l},\label{SC}
\end{eqnarray}
where $J_i=J_0/\Delta_i^\beta$, $\hat{c}_{j, l}$ satisfying $[\hat{c}_{j,l},\hat{c}_{j',l'}^\dag]_+=\delta_{ll'}\delta_{jj'}$ is the fermionic annihilation operator of the $j$th sublattice of the $l$th unit cell, and a constant has been abandoned. The nearest-neighbor interaction approximation is justified as follows. First, we can make the nearest-neighbor interaction dominant by adjusting the collective vibration modes of ions. The ions generally have the vibrational modes of the center of mass, the tilt, and the zigzag. Contributing different types of spin-spin interaction, they can be controlled by the laser detuning. If the ions are in the zigzag mode, then the nearest-neighbor interaction is dominant \cite{Kim_2011}. Second, the nearest-neighbor interaction tends to dominate at larger spatial separation on the ion chain \cite{PhysRevA.90.053405}. Third, the longer-range interaction gives less contributions when the power index $\beta$ is large \cite{PRXQuantum.4.010302,PhysRevA.100.032115}. Thus, we can reasonably consider only the nearest-neighbor interaction for $\beta$ being its largest value, i.e., three \cite{Vodola_2016}.

Equation \eqref{SC} hosts a $p$-wave topological-superconductor-like phase. To reveal its bulk-boundary correspondence, we rewrite Eq. \eqref{SC} in the momentum space under the periodic-boundary condition as $\hat{H}=\sum_{k}\hat{\mathbf C}^{\dagger}_{k}\mathcal{H}(k)\hat{\mathbf C}_{k}$ with $\hat{\textbf{C}}^{\dagger}_{k} = (\hat{\tilde c}^{\dagger}_{a, k},\hat{\tilde c}_{a, -k}, \hat{\tilde c}^{\dagger}_{b, k},\hat{\tilde c}_{b, -k}$), where $\hat{\tilde c}_{j,k}=\sum_l\hat{c}_{j,l}\exp(ikl)/\sqrt{N/2}$. The Bogoliubov-de Gennes Hamiltonian reads
\begin{eqnarray} 
\mathcal{H}(k)&=&[2B\tau_{0}+(J_1+J_2\cos k)\tau_{x}+J_2\sin k\tau_{y}]s_{z}\nonumber\\
&&-[J_2\sin k\tau_{x}+(J_1-J_2\cos k)\tau_{y}]s_{y},\label{SCK1}
\end{eqnarray}
where $\tau_{i}$ and $s_{i}$, respectively, are the Pauli matrices acting on the sublattice and particle-hole subspaces, and $\tau_{0}$ is the identity matrix. $\mathcal{H}(k)$ has particle-hole ${\mathcal{C}=\tau_0s_xK}$, time-reversal ${\mathcal{T}=K}$, with $K$ being the complex conjugation, and chiral ${\mathcal{S}=\tau_0s_x}$ symmetries. Thus, it belongs to the topological class BDI and its bulk-band topology is characterized by the winding number \cite{PhysRevB.78.195125}. Equation \eqref{SCK1} is unitarily equivalent to an antidiagonal matrix $\left(
  \begin{array}{cc}
    0 & D(k) \\
    D^\dag(k) & 0 \\
  \end{array}
\right)$, with $D(k)=2[J_2\sin k+i(J_1-J_2\cos k)]\tau_x-2[iJ_2\sin k+(J_1+J_2\cos k)]\tau_{y}+4iB\tau_{z}$. The winding number for our four-band system is defined as $W=\int^{\pi}_{-\pi}\frac{dk}{2\pi i} \partial_k\text{ln}[\text{det} D(k)]$, which denotes the number of the Majorana modes with zero energy \cite{doi:10.1126/science.1259327,RN222}. It is remarkable to find that the system also has the Majorana modes with nonzero energies. To characterize the topological features of such nonzero-energy Majorana modes, we resort to the dipole moment $P=\big[\frac{\text{Im}\ln\det\mathcal{F}}{2\pi}-\sum_{j,l;j',l'} \frac{r_{j,l;j',l'}}{4N}\big]\text{mod}~1$, where $\mathcal{F}_{pp'}\equiv \langle \psi_p|e^{i2\pi r/N}|\psi_{p'}\rangle$, with $|\psi_p\rangle$ satisfying $\hat{H}|\psi_p \rangle=E_{p}|\psi_p\rangle$ are the lowest occupied eigenstates, and the coordinate $r_{j,l;j',l'}=l\delta_{jj'}\delta_{ll'}$ with $j$ and $j'$ being the sublattice index and $l$ and $l'$ being the unit-cell number \cite{PhysRevA.106.013305,PhysRevB.100.245134,PhysRevB.100.245135}. Physically, $P$ signifies the formation
of the topological phases via describing the density distribution of the
relevant fermions in the occupied states of the system \cite{PhysRevA.106.013305,doi:10.1126/science.aah6442}. It is readily calculated from Eq. \eqref{SCK1} that the upper two bands close at a nonzero energy when $|J_1|=|J_2|$, where the phase transition characterized by $P$ occurs, and the middle two bands close at the zero energy when 
\begin{equation}
    2B^2+J_1^2+J_2^2=\sqrt{4B^2(J_1+J_2)^2+(J_1^2-J_2^2)^2},\label{btcd}
\end{equation}
where the phase transition characterized by $W$ occurs. 

In the uniform case of $\Delta_1=\Delta_2$, we have $J_1=J_2\equiv J$. The system reduces to a two-band model and the phase transition characterized by $P$ does not occur. It is calculated that $D(k)=
2i(J\tau_+-Je^{ik}\tau_-+2B\tau_z)$, with $\tau_\pm=(\tau_x\pm i\tau_y)/2$, and thus $W=1$ for $|J|>|B|$ and $0$ for $|J|<|B|$. The energy spectrum under the open-boundary condition in Fig. \ref{fig:1}(b) confirms that the zero-energy Majorana modes are present in the regime of $W=1$. They are twofold degenerate and distribute at the two lattice edges, respectively, see Fig. \ref{fig:1}(c). The phase diagram described by $W$ in Fig. \ref{fig:1}(d) gives a global picture on the topological phases, whose boundaries agree to the band-coalesce condition $|J|=|B|$ at the zero energy obtained from Eq. \eqref{btcd}. 

\begin{figure}[tbp]
\centering
\includegraphics[width=\columnwidth]{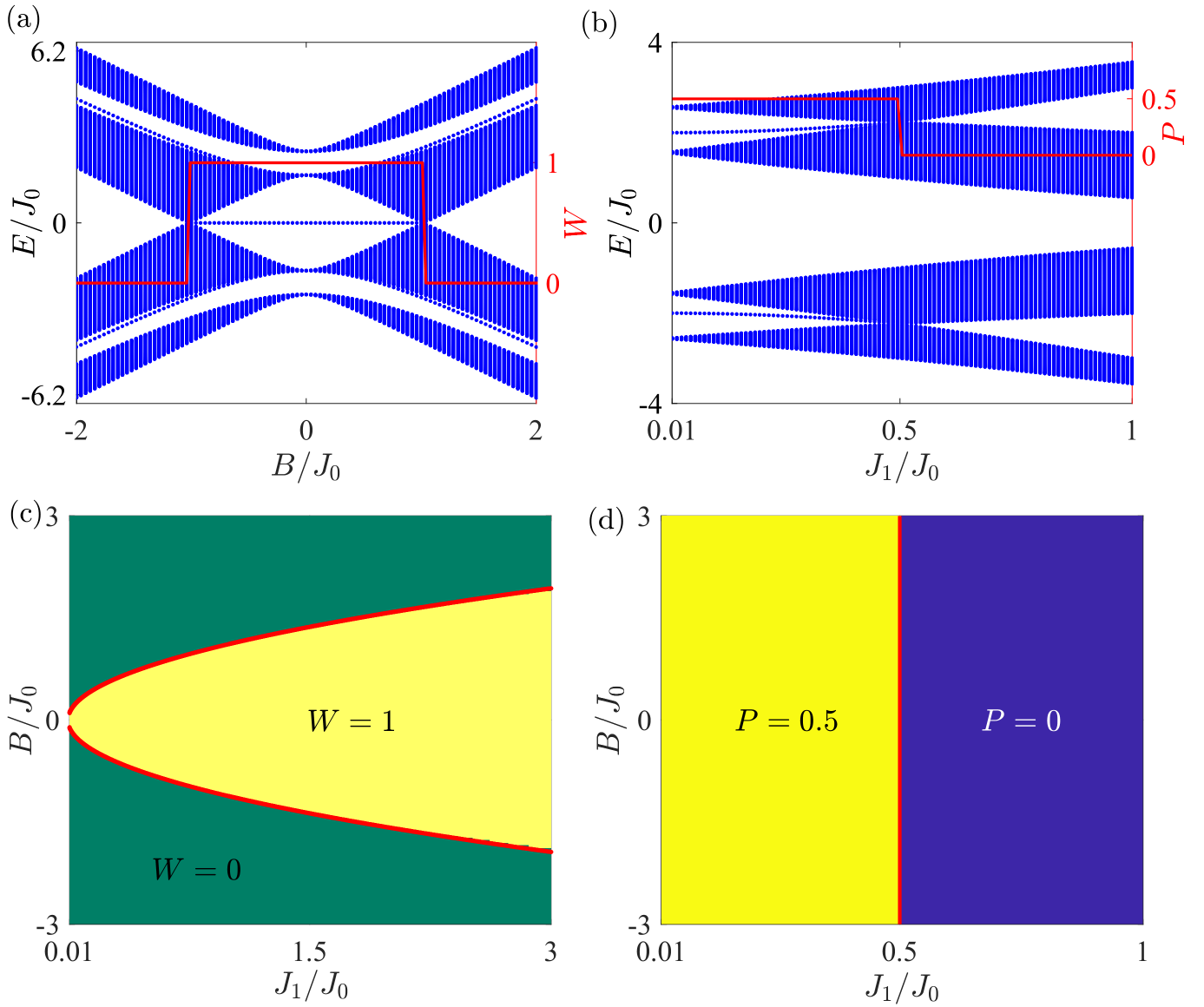}
\caption{(a) Energy spectrum and winding number $W$ (red line) in different $B$ when $J_1=5J_0/6$ and $J_2=5J_0/4$. (b) Energy spectrum and dipole moment $P$ (red line) in different $J_1$ when $B=J_0$ and $J_2=0.5J_0$. (c) Zero-mode phase diagram described by $W$ when $J_2=5J_0/4$. (d) Nonzero-mode phase diagram described by $P$ when $J_2=0.5J_0$.}\label{fig:2}
\end{figure}

The energy spectrum in the nonuniform case of $\Delta_1\neq \Delta_2$ shows that the system hosts the formation of the Majorana modes not only at the zero energy but also at the nonzero energies, see Figs. \ref{fig:2}(a) and \ref{fig:2}(b). The zero-energy Majorana modes present when $W=1$ and the nonzero-energy ones present when $P=0.5$. Figure \ref{fig:2}(c) shows the phase diagram of the zero-energy Majorana modes described by $W$ in the $B$-$J_1$ plane, where the phase boundaries obey Eq. \eqref{btcd}. Figure \ref{fig:2}(d) shows the nonzero-energy one described by $P$, where the phase transition occurs at $|J_1|=|J_2|$. The results indicate that our proposed trapped-ion system possesses rich Majorana modes. Our dimerized-ion configuration with the formation of the Majorana modes not only at the zero but also the nonzero energy gaps enhances the Majorana modes, which could be used to encode more qubits in quantum computing by the braiding of Majorana modes.

\section{Floquet engineering}
For quantum computing, how to efficiently generate and annihilate different numbers of Majorana modes is an important question. Limited by the finite control methods, it is usually difficult in static systems. We propose to control the Majorana modes by Floquet engineering. 

First, we consider that the periodic driving is applied on the hopping rate $J$ in the uniform case as
\begin{equation}\label{FE}
\begin{split}
J(t)= \left \{
 \begin{array}{ll}
U_1,                    & t\in [nT,nT+T_{1})\\
U_2,                    & t\in [nT+T_{1},(n+1)T)
 \end{array}
 \right.,
 \end{split}
 \end{equation}
where $n\in\mathbb{Z}$ and $T = T_{1} +T_{2}$ is the driving period. This may be realized by periodically manipulating either the ion separation between two spatial configurations or the Rabi frequency. The periodic system $\hat{H}(t)$ does not have a well-defined energy spectrum because its energy is not conserved. According to the Floquet theorem, we can define an effective Hamiltonian $\hat{H}_{\text{eff}}=\frac{i}{T}\ln{\hat{\mathcal U}_{T}}$ from one-period evolution operator $\hat{\mathcal U}_{T} = \mathbb{T}e^{-i\int^{T}_{0} \hat{H}(t) dt}$, with $\mathbb{T}$ being the time-ordering operator. The eigenvalues of $\hat{H}_{\text{eff}}$ are called quasienergies and the topological properties of the periodic system are defined in the quasienergy spectrum \cite{PhysRevA.7.2203,PhysRevB.96.195303}. Applying the Floquet theorem to our system, we obtain $\mathcal{H}_{\text{eff}}(k)=\frac{i}{T}\ln[\text{e}^{-i\mathcal{H}_{2}(k)T_{2}}\text{e}^{-i\mathcal{H}_{1}(k)T_{1}}]$, where $\mathcal{H}_{j}(k)$ is Eq. \eqref{SCK1} with $J_1=J_2\equiv J$ replaced by $U_j$.

Periodic systems have unique $\pi/T$-quasienergy topological phases, which, although enrich the bulk-boundary correspondence, make the topological descriptions well defined in static fermionic systems inadequate. To reveal the complete bulk-boundary correspondence of our periodic fermionic system, proper topological invariants to describe both the zero- and $\pi/T$-quasienergy Majorana modes are needed. Unfortunately, the chiral symmetry of $\mathcal{H}_j(k)$ is not inherited by $\mathcal{H}_\text{eff}(k)$ due to $[\mathcal{H}_{1}({k}), \mathcal{H}_{2}({k})]\neq 0$, which means that we cannot define a winding number in $\mathcal{H}_\text{eff}(k)$ in a similar manner as the static case. To restore chiral symmetry, we make two unitary transformations $G_{l}(k)=e^{i(-1)^{l}\mathcal{H}_{l}(k)T_{l}/2}$ ($l=1,2$) to $\mathcal{H}_\text{eff}(k)$, which do not change the quasienergy spectrum, and obtain two chirally symmetric $\mathcal{\widetilde{H}}_{\text{eff},l}(k) = iT^{-1}\ln[{G_{l}(k){\mathcal U}_{T}(k)G_{l}^{\dag}(k)}]$. Then, two winding numbers $W_l$ are well defined in $\mathcal{\widetilde{H}}_{\text{eff},l}(k)$ and the topological features at the quasienergies $\alpha/T$, with $\alpha=0$ or $\pi$, are described by $W_{\alpha/T}=(W_{1}+e^{i\alpha}W_{2})/2$. The number of $\alpha/T$-quasienergy Majorana modes equals to $2|W_{\alpha/T}|$ \cite{PhysRevB.90.125143}. The phase boundaries also can be obtained from $\mathcal{H}_{\text{eff}}(k)$. We find that the topological phase transition occurs at the quasienergies $\alpha/T$ for the system and driving parameters making the eigenvalues of $\hat{\mathcal U}_{T}$ be $e^{i\alpha}$. Thus, we can readily obtain the phase boundaries as either
\begin{equation}\label{btts1}
\sqrt{4U_j^2-8BU_j\cos k+4B^2}T_j=n_j\pi,
\end{equation}
or 
\begin{equation}\label{btts2}
|U_1+Be^{i\gamma}|T_1 \pm |U_2+Be^{i\gamma}|T_2=n_{\gamma,\pm}\pi/2,
\end{equation}
with $\gamma=0$ and $\pi$, see Appendix \ref{appptc}, at the quasienergy zero (or $\pi/T$) when $n_1$ and $n_2$ in Eq. \eqref{btts1} are integers with a same (or different) parity and $n_{\gamma,\pm}$ in Eq. \eqref{btts2} is even (or odd).

\begin{figure}[tbp]
\centering
\includegraphics[width=\columnwidth]{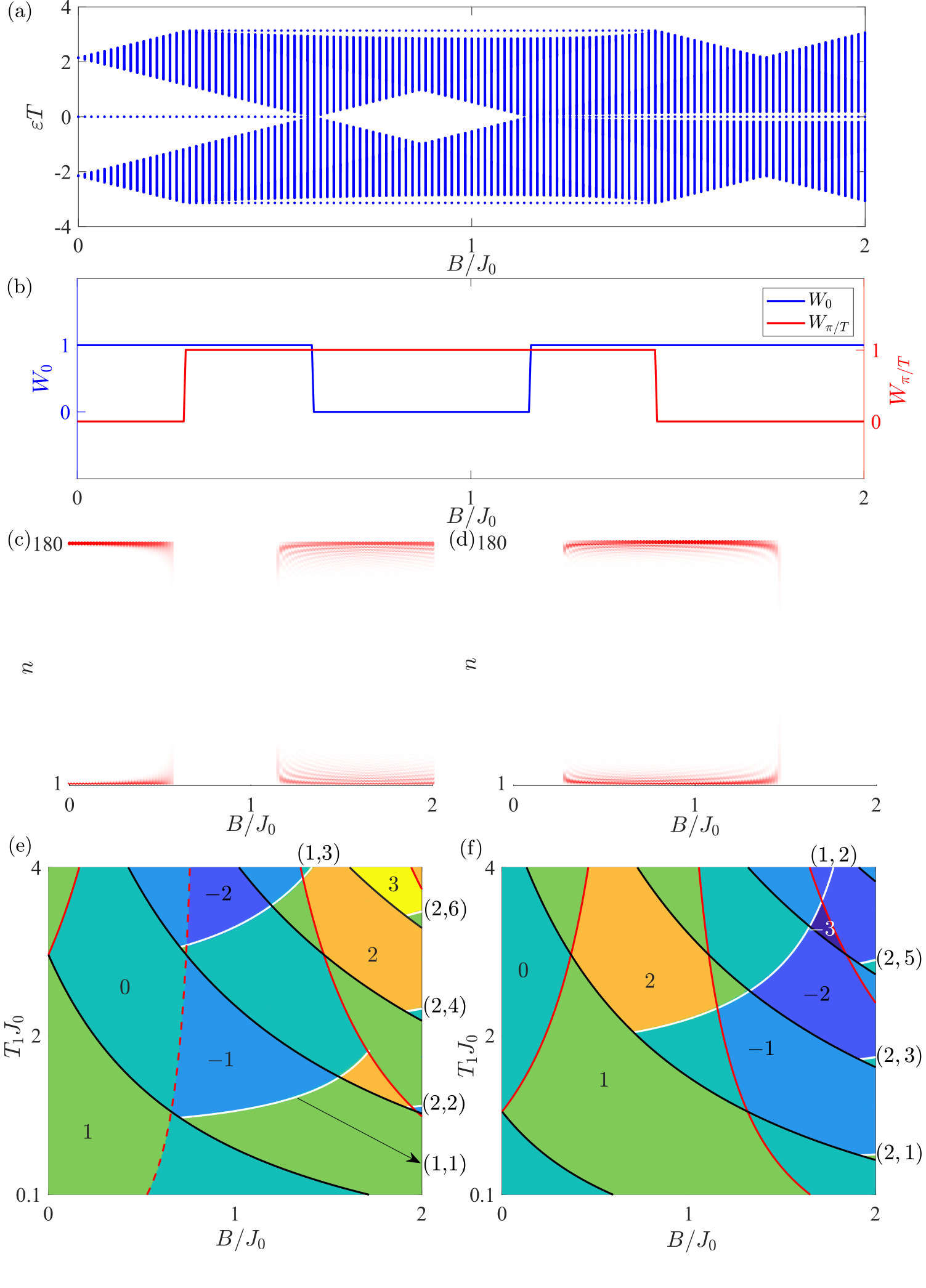}
\caption{(a) Quasienergy spectrum and (b) winding numbers $W_{\alpha/T}$ of the periodic system in different $B$ at $T_1=0.5J_0^{-1}$. Probability distributions of the (c) zero- and (d) $\pi/T$-quasienergy Majorana modes. Phase diagram described by (e) $W_0$ and (f) $W_{\pi/T}$. The white solid lines are from Eq. \eqref{btts1} with the labeled ($n_1,n_2$). Equation \eqref{btts2} with $n_{0,+}=2, 4$ in (e) and $n_{0,+}=1,3$ in (f) is depicted by red solid lines, with $n_{0,-}=0$ by the red dashed line and with $n_{\pi,+}=2,4,6,8$ by the black solid lines in (e), and $n_{\pi,+}=1,3,5,7,9$ by the black solid lines in (f). We use $U_1=0.8J_0$, $U_2=4J_0/3$, and $T_2=1.3J_0^{-1}$.}\label{fig:3}
\end{figure}

Figure \ref{fig:3}(a) shows the quasienergy spectrum in the open-boundary condition, whose topological features are well described by the winding number $W_{\alpha/T}$ in Fig. \ref{fig:3}(b). It is interesting to see that the coexisting Majorana modes in both the quasienergies zero and $\pi/T$ are present. Both of the two types of Majorana modes possess the feature of lattice-edge distribution, see Figs. \ref{fig:3}(c) and \ref{fig:3}(d). The phase diagrams characterized by $W_{\alpha/T}$ in Fig. \ref{fig:3}(e) and Fig. \ref{fig:3}(f) reveal that, in contrast to the static case in Fig. \ref{fig:1}(d), the periodic system has a widely tunable $W_{\alpha/T}$ from $-3$ to $3$, whose phase boundaries match well with our analytic result in Eqs. \eqref{btts1} and \eqref{btts2}. It means that we can freely manipulate the number of the Majorana modes by changing the driving parameters. The distinguished role played by the periodic driving of Floquet engineering in generating rich Majorana modes is that it can induce effective long-range hopping of the system among the lattice sites, which efficiently fold the (quasi)energy bands \cite{PhysRevB.87.201109} and creates multiple band-touching points \cite{PhysRevB.93.184306}. Consequently, an amount of topological phases absent in the static system, including the large winding number phases, are generated. Thus, Floquet engineering supplies us with a useful tool in controlling the Majorana modes in our trapped-ion system.

\begin{figure}[tbp]
\centering
\includegraphics[width=\columnwidth]{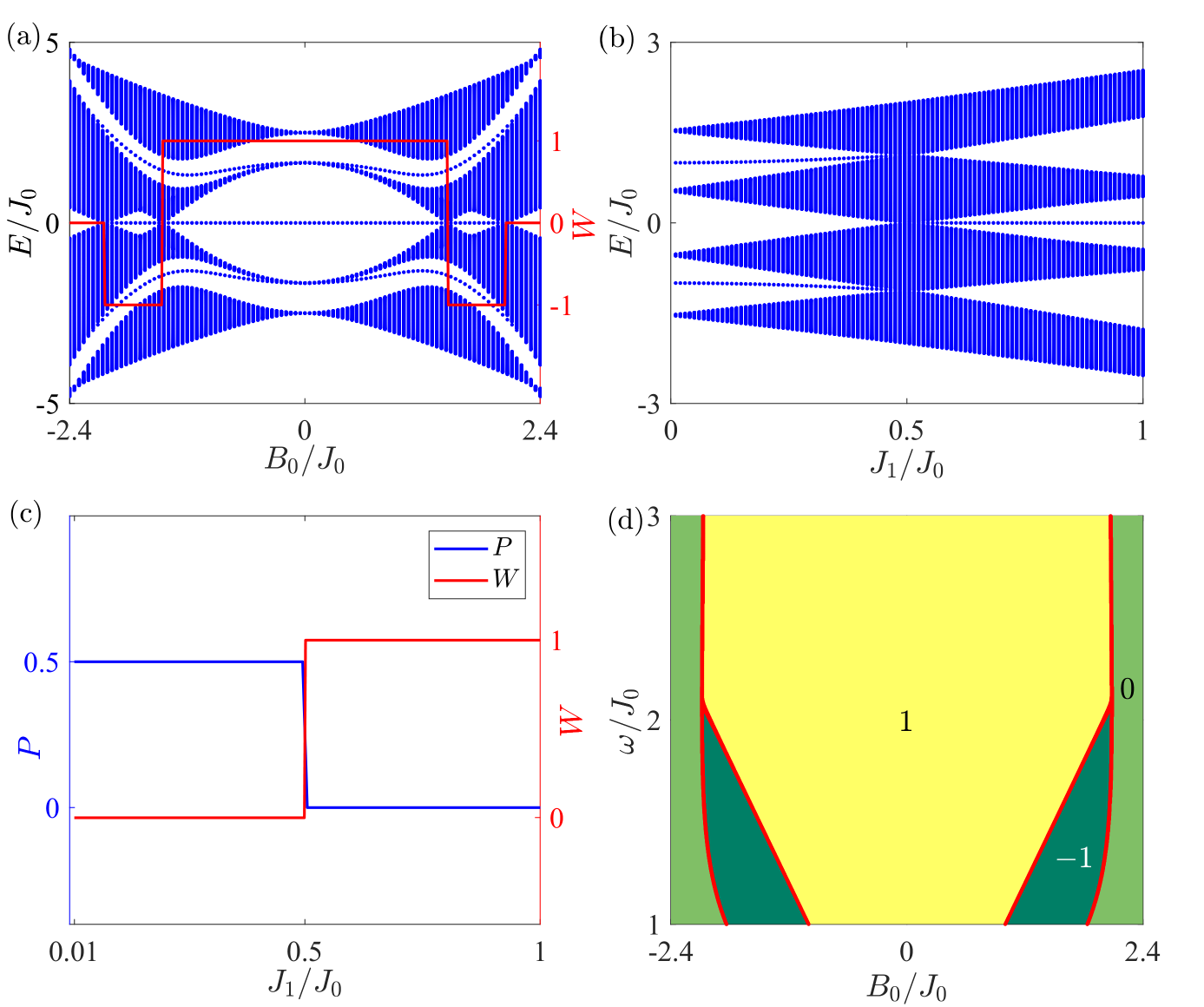}
\caption{(a) Quasienergy spectrum and winding number $W$ (red line) in different $B_0$ when $J_1=5J_0/6$ and $J_2=5J_0/4$. (b) Quasinergy spectrum and (c) dipole moment $P$ (red line) and winding number $W$ (blue line) in different $J_1$ when $\omega=1.45J_0$, $B=J_0$, and $J_2=5J_0/4$. (d) Phase diagram described by $W$ when $J_1=5J_0/6$ and $J_2=5J_0/4$.}\label{fig:4}
\end{figure}

Second, we apply Floquet engineering to the nonuniform case in Eq. \eqref{SCK1}. We consider that the transverse field is periodically changed as $B(t)=B_0\sin^2(\omega t)$, with $\omega$ being the driving frequency. This can be realized by manipulating the two Raman-field beatnotes. In the high-frequency limit, we can make a second-order Magnus expansion to the one-period evolution operator and obtain \cite{doi:10.1080/00018732.2015.1055918,Eckardt_2015}
\begin{eqnarray}
&&\mathcal{H}_{\text{eff}}(k)\simeq [B_0\tau_{0}+(J_1+J_2\cos k)\tau_{x}+J_2\sin k\tau_{y}]s_{z}\nonumber\\
&&~~~~~~-(1-{B_0^2\over\omega^2})[J_2\sin k\tau_{x}+(J_1-J_2\cos k)\tau_{y}]s_{y}.\label{effhmt}
\end{eqnarray} 
Equation \eqref{effhmt} possesses the same symmetries as Eq. \eqref{SCK1}. Its topological feature can be described by the similar method to the static case. Its hopping amplitude along $s_y$ is renormalized by the driving parameters. This gives a sufficient space to control the topological phase transition by adjusting the external periodic-driving field. It is easy to derive that the nonzero-quasienergy topological phase transition still occurs at $|J_1|=|J_2|$, while the zero-quasienergy one occurs at 
\begin{eqnarray}
&&B_0^2+(1+f^2)(J_1^2+J_2^2)+2(1-f^2)J_1J_2\cos k=\nonumber\\
&&2\sqrt{B_0^2(J_1^2+2J_1J_2\cos k+J_2^2)+f^2(J_1^2-J_2^2)^2},
\end{eqnarray}
where $f=1-B_0^2/\omega^2$.
Figures \ref{fig:4}(a) and \ref{fig:4}(b) show the quasienergy spectrum of Eq. \eqref{effhmt} under the open-boundary condition. The topological phase transition at both the zero and nonzero gaps of the quasienergies are well described by $W$ and $P$, see Fig. \ref{fig:4}(c). We find that the periodic driving can create the Majorana modes from the topologically trivial static fermionic system in Figs. \ref{fig:2}(a) and \ref{fig:2}(b). The phase diagram in Fig. \ref{fig:4}(d) also clearly proves that Floquet engineering can create the Majorana modes in the regions where the static system does not host their existence. 

\section{Discussion and conclusion}
Our dimerization approach to implement the Majorana modes can be further extended to trimerization and so on, where the number of Majorana modes could be further increased. The topologically trivial states correspond to the paramagnetic states in the spin model. The topological states in the presence of the Majorana modes correspond to the ferromagnetic states polarized in either $+x$ or $-x$ in the spin model. The paramagnetic states and the ferromagnetic states may be detected by measuring the magnetic moment $M_x=\langle\sum_i\hat{\sigma}^x_i\rangle$.
The Majorana modes with nonzero energies can also be detected by measuring the dipole moment $P$. The recent experimental progress on trapped-ion systems supports the realization of our proposal \cite{PhysRevLett.130.163001,PRXQuantum.4.010302,PhysRevA.108.012415,RN223,RN224,RN225}. Furthermore, although our proposal is based on the trapped-ion system, it is applicable to the Rydberg-atom array hold by optical tweezers, where the Ising model has been experimentally realized \cite{PhysRevLett.120.113602,RN213,RN216,RN215,RN220,RN221}. Our scheme only reveals a minimal requirement to simulate Majorana modes in trapped-ion systems under the nearest-neighbor approximation. It is interesting to further explore the correction of the longer-range interactions to our leading-order result. According to Ref. \cite{Vodola_2016}, even in the presence of the longer-range interactions, the main conclusion on forming the Majorana modes still does not qualitatively change. A final remark is that the nonlocal Jordan-Wigner transformation might trigger
a dichotomy of a formal equivalence and a physical inequivalence between the original Ising model and the Kitaev-chain model such that the ground state in the latter is topologically protected while in the former it is not \cite{GREITER20141026}. However, the system still can be used to perform quantum computing via the braiding of the topologically nonprotected ferromagnetic states in the Ising model \cite{PhysRevB.96.195402}. 

In summary, we propose a scheme to realize Majorana modes and its Floquet engineering in a trapped-ion system. Our dimerized-ion configuration supports the formation of the Majorana modes not only at the zero, but also the nonzero energy gaps. We also propose to create and annihilate different numbers of Majorana modes by Floquet engineering. A widely tunable number of Majorana modes can be created on demand in the static topologically trivial fermionic system by applying periodic driving. Providing an alternative platform to controllably simulate the mysterious Majorana fermions, our scheme paves the way to explore quantum computing by the braiding of Majorana modes in trapped-ion systems.

\section*{Acknowledgements}
The work is supported by the National Natural Science Foundation (Grants No. 12275109 and No. 12247101).

\appendix
\section{System}\label{appsys}
The evolution operator of Eq. \eqref{ham1} is
\begin{equation}
\hat{U}(t)=\mathbb{T}e^{-i\int^{t}_{0} \hat{ H}(t') dt'},\label{E1}
\end{equation} 
where $\mathbb{T}$ is the time-ordering operator. According to the Magnus formula $\hat{U}(t)=\exp\{-i\int^{t}_{0} \hat{H}(t') dt'-\frac{1}{2}\int^{t}_{0}dt_2\int^{t_2}_{0}[\hat{H}(t_2),\hat{H}(t_1)]dt_1+\cdots\}$, Eq. \eqref{E1} is expanded as \cite{PhysRevLett.91.187902,PhysRevLett.97.050505}
\begin{equation}
\hat{U}(t)=\exp\Big[\sum_{j=1}^{N}\phi_j(t)\hat{\sigma}_j^x+\sum_{p,q=1}^N\chi_{p,q}(t)\hat{\sigma}_p^x\hat{\sigma}_q^x\Big],\label{E2}
\end{equation} 
where $\phi_j(t)=\sum_m[g_{j,m}(t)\hat{a}_m^\dag-g_{j,m}^*(t)\hat{a}_m]$. The first term is spin-dependent displacements of the $m$th phonon mode by an amount
\begin{equation}
 g_{j,m}(t)=\frac{-i\eta_{j,m}\Omega}{\mu^2-\omega_m^2}[\mu-e^{i\omega_mt}(\mu\cos\mu t-i\omega_m\sin\mu t)\big].
\end{equation}
The second term is a spin-spin interaction between the $p$th and $q$th ions with coupling strength 
\begin{eqnarray}
  \chi_{p,q}(t)&=&\frac{\Omega^2}{2}\sum_m\frac{i\eta_{p,m}\eta_{q,m}}{\mu^2-\omega_m^2}\big[\frac{\mu\sin(\mu-\omega_m)t}{\mu-\omega_m}\nonumber\\
  &&-\frac{\mu\sin(\mu+\omega_m)t}{\mu+\omega_m}+\frac{\omega_m\sin 2\mu t}{2\mu}-\omega_mt\big]. ~~\label{smchi}
\end{eqnarray} We focus on the ``slow'' regime, where the optical beatnote frequency is far detuning from one of each normal mode, i.e., $|\mu-\omega_m|\gg\Omega\eta_{j,m}$. Then, the phonons are only virtually excited and the ion displacements become negligible, i.e., $\phi_j(t)\simeq 0$ \cite{RN194,PhysRevLett.103.120502}. Under the rotating-wave approximation, only the last term of Eq. \eqref{smchi} is kept. In this case, Eq. \eqref{E2} represents the dynamics of the pure Ising model $\hat{H}=\sum_{i,j}J_{i,j}\hat{\sigma}_i^x\hat{\sigma}_j^x$ with
\begin{equation}
J_{i,j}=\frac{\Omega^2(\delta k)^2}{4M}\sum_m\frac{b_{p,m}b_{q,m}}{\mu^2-\omega_m^2}.\label{cp}
\end{equation} 
Further, we adjust the two Raman beatnotes to $\omega_0\pm\mu+B$. A uniform effective transverse magnetic field of $B$ along $\hat{\sigma}_i^z$ is generated \cite{RN194}. The coupling strength $J_{ij}$ is approximated as a power law $J_{i,j}\simeq J_0/|z_i-z_j|^\beta$, with $\beta\in (0,3)$ \cite{RN197,RN194,RN195,doi:10.1126/science.1232296,PhysRevLett.92.207901}.

In order to analyze the properties of the system, we take the Jordan-Wigner transformation \cite{PhysRevLett.115.177204} 
\begin{eqnarray}
\hat{\sigma} _{j}^{z} &=&1-2\hat{c}_{j}^{\dagger }\hat{c}_{j}\text{, }\hat{\sigma} _{j}^{y}=\mathrm{i
}\hat{\sigma} _{j}^{x}\hat{\sigma }_{j}^{z}, \nonumber\\
\hat{\sigma} _{j}^{x} &=&-\prod\limits_{l<j}\big( 1-2\hat{c}_{l}^{\dagger }\hat{c}_{l}\big)
\big( \hat{c}_{j}+\hat{c}_{j}^{\dag }\big) ,  \label{JW}
\end{eqnarray}
where $\hat{c}_{j}^{\dagger }$ is the fermionic generation operator. We propose that the ion array has a dimerization configuration, which can be realized by setting the distance between each odd (even) ion and its next neighboring ion being $\Delta_1$ ($\Delta_2$). Thus, the ion array reduces into a lattice of $N/2$ unit cells, each of which contains two sublattices labeled by $a$ and $b$. The fermionized Hamiltonian reads
\begin{eqnarray} 
\hat{H}&=&\sum_{l=1}^{N/2-1}[J_1\hat{c}_{a,l}^{\dagger}(\hat{c}_{b,l}+\hat{c}_{b,l}^{\dagger})+J_2\hat{c}_{b,l}^{\dagger}(\hat{c}_{a,l+1}+\hat{c}_{a,l+1}^{\dagger})\nonumber\\
&&+\text{H.c.}]-2B\sum_{j=a,b}\sum_{l=1}^{N/2}\hat{c}_{\alpha,l}^{\dagger}\hat{c}_{\alpha,l},\label{SCap}
\end{eqnarray}
where $J_i=J_0/\Delta_i^\beta$, $\hat{c}_{j, l}$ satisfying $[\hat{c}_{j,l},\hat{c}_{j',l'}^\dag]_+=\delta_{ll'}\delta_{jj'}$ is the fermionic annihilation operator of the $j$th sublattice of the $l$th unit cell, and a constant has been abandoned.

In the homogeneous case where $\Delta_1=\Delta_2$, the ion array becomes a uniform array and $J_1=J_2\equiv J$. Thus, the degrees of freedom of the sublattice, i.e., $a$ and $b$, can be absorbed. Equation \eqref{SCap} reduces to
\begin{eqnarray} 
\hat{H}=\sum_{l=1}^{N-1}[J\hat{c}_{l}^{\dagger}(\hat{c}_{l+1}+\hat{c}_{l+1}^{\dagger})+\text{H.c.}]-2B\sum_{l=1}^{N}\hat{c}_{l}^{\dagger}\hat{c}_{l}.\label{SC1}
\end{eqnarray}

\section{Topological phase transition induced by periodic driving}\label{appptc}
We can derive an effective Hamiltonian $\hat{H}_{\text{eff}}=\frac{i}{T}\ln\hat{\mathcal U}_{T}$ from one-period evolution operator $ \hat{\mathcal U}_{T} = \mathbb{T}e^{-i\int^{T}_{0} \hat{H}(t) dt}$ for a periodically driven system. Its topological properties are defined in the eigenvalues of $\hat{H}_{\text{eff}}$. After making the Fourier transform under the periodic boundary condition, Eq. \eqref{SC1} is rewritten as $\hat{H}=\sum_{k}\hat{\mathbf C}^{\dagger}_{k}\mathcal{H}(k)\hat{\mathbf C}_{k}$ with $\hat{\textbf{C}}^{\dagger}_{k} = (\hat{\tilde c}^{\dagger}_{k},\hat{\tilde c}_{-k}$), where $\hat{\tilde c}_{k}=\sum_l\hat{c}_{l}\exp(ikl)/\sqrt{N}$. The Bogoliubov-de Gennes Hamiltonian reads 
\begin{equation} \label{SCK}
\mathcal{H}(k)=(2J\cos k-2B)\tau_{z} -2J\sin k \tau_{y}\equiv \mathbf{d}(k)\cdot \pmb\tau.
\end{equation}
Under the driving protocol in Eq. \eqref{FE} and using Euler’s formula of the Pauli matrices, we obtain the one-period evolution operator as
\begin{eqnarray}
\mathcal U_{T} &=&e^{-i\mathbf{d}_{2}(k)\cdot \pmb\tau T_{2}}e^{-i\mathbf{d}_{1}(k)\cdot \pmb\tau T_{1}}= \epsilon I_{2\times 2}-i{\bf r}\cdot{\pmb\tau},
\end{eqnarray}
where $ \epsilon$ and ${\bf r}$ are
\begin{eqnarray}
 \epsilon &=&\cos (d_{1}T_{1})\cos (d_{2}T_{2})-\sin (d_{1}T_{1})\sin (d_{2}T_{2})%
\underline{\mathbf{d}}_{1}\cdot \underline{\mathbf{d}}_{2}, \nonumber\\
{\bf r} &=&\underline{\mathbf{d}}_{1}\sin (d_{1}T_{1})\cos (d_{2}T_{2})+\underline{\mathbf{d}}_{2}\cos (d_{1}T_{1})\sin (d_{2}T_{2})\nonumber\\
&-&\sin (d_{1}T_{1})\sin (d_{2}T_{2})\underline{\mathbf{d}}_{1}\times \underline{\mathbf{d}}_{2},\label{PQ}
\end{eqnarray}with ${\mathbf d}_j(k)=d_j\underline{\mathbf d}_j$.
The unitariness of $\mathcal U_{T}$ requires $\epsilon^{2}+|{\bf r}|^{2}=1$. Thus the effective Hamiltonian is
\begin{equation}
\mathcal H_{\text{eff}}(k)=\frac{\arccos \epsilon}{T}\frac{{\bf r}\cdot{\pmb\tau}}{|{\bf r}|}.
\end{equation}%
The eigenvalues of $\mathcal{H}_{\text{eff}}(k)$ are $\varepsilon =\pm \frac{\arccos \epsilon}{T}$, which are the quasienergies. We find that the topological phase transition occurs at the quasienergies zero and $\pi/T$ when $\epsilon=1$ and $-1$, respectively. According to Eq. \eqref{PQ}, the phase transition occurs for $k$ and driving parameters satisfying one of the following:
\begin{enumerate}
  \item $\sin(d_1T_1)\sin(d_2T_2)=0$. In this case, $\epsilon=\cos(d_1T_1)\cos(d_2T_2)$. Then the bands of $\mathcal{H}_\text{eff}(k)$ close when
  \begin{equation}
d_jT_j=n_j\pi,~n_j\in \mathbb{Z} \label{bt1}
  \end{equation}
at the quasienergy zero (or $\pi/T$) if $n_{1}$ and $n_{2}$ are integers with same (different) parity.
  \item $\underline{\mathbf{d}}_1\cdot\underline{\mathbf{d}}_2=\pm1$. In this case, $\epsilon=\cos(d_1T_1\pm d_2T_2)$. Then the bands of $\mathcal{H}_\text{eff}(k)$ close when
  \begin{equation}
  d_1T_1\pm d_2T_2=n\pi,~n\in\mathbb{Z}\label{bt2}
  \end{equation}
 at the quasienergy zero (or $\pi/T$) if $n$ is even (or odd).
\end{enumerate}
By combining Eqs. \eqref{SCK}, \eqref{bt1}, and \eqref{bt2}, we can get the phase boundaries satisfying either
\begin{equation}
\sqrt{4U_j^2-8BU_j\cos k+4B^2}T_j=n_j\pi
\end{equation}
or 
\begin{equation}
|U_1+Be^{i\gamma}|T_1 \pm |U_2+Be^{i\gamma}|T_2=n_{\gamma,\pm}\pi/2,
\end{equation}
with $\gamma=0$ and $\pi$.
\bibliography{references}
\end{document}